\documentclass[12pt]{article}

\usepackage{epsfig,amsmath,amssymb,latexsym,axodraw}

\providecommand{\beqa}{\begin{eqnarray}}
\providecommand{\eeqa}{\end{eqnarray}}

\providecommand{\om}{{\omega}}
\providecommand{\Om}{{\Omega}}
\providecommand{\alphabar}{{\bar{\alpha}}}
\providecommand{\betabar}{{\bar{\beta}}}

\providecommand{\add}{{\ddot a}}
\providecommand{\Ombar}{\overline{\Omega}}
\providecommand{\jbar}{\bar{\jmath}}

\providecommand{\Ibar}{\overline{I}}
\providecommand{\Jbar}{\overline{J}}

\providecommand{\Sbar}{\overline{S}}
\providecommand{\Tbar}{\overline{T}}
\providecommand{\Ybar}{\overline{Y}}
\providecommand{\Wbar}{\overline{W}}

\providecommand{\we}{\wedge}

\def\cM{{\cal{M}} }
\def\cO{{\cal{O}} }
\def\cP{{\cal{P}} }
\def\cR{{\cal{R}} }
\def\Orb{{\mathbf{S}^1/\mathbf{Z}_2} }
\def\cV{{\cal{V}}}

\numberwithin{equation}{section}

\begin{document}

\thispagestyle{empty} \phantom{}\vspace{-1.5cm}
\rightline{UMD-PP-05-032}

\begin{center}
{\bf \Large M-Theory Inflation from Multi M5-Brane Dynamics}
\end{center}

\vspace{0.6truecm}

\centerline{Katrin Becker$^{a,}$\footnote{katrin@physics.utah.edu},
Melanie Becker$^{b,}$\footnote{melanieb@physics.umd.edu} and Axel
Krause $^{b,}$\footnote{krause@physics.umd.edu}}

\vspace{.6truecm}

{\em \centerline{$^a$ Department of Physics, University of Utah,}
\centerline{Salt Lake City, UT 84112-0830, USA}}

\vspace{.3truecm}

{\em \centerline{$^{b}$ Department of Physics, University of
Maryland,} \centerline{College Park, MD 20742, USA}}

\vspace{0.6truecm}

\begin{abstract}
We derive inflation from M-theory on $\Orb$ via the
non-perturbative dynamics of $N$ M5-branes. The open membrane
instanton interactions between the M5-branes give rise to
exponential potentials which are too steep for inflation
individually but lead to inflation when combined together. The
resulting type of inflation, known as assisted inflation,
facilitates considerably the requirement of having all moduli,
except the inflaton, stabilized at the beginning of inflation.
During inflation the distances between the M5-branes, which
correspond to the inflatons, grow until they reach the size of the
$\Orb$ orbifold. At this stage the M5-branes will reheat the
universe by dissolving into the boundaries through small instanton
transitions. Further flux and non-perturbative contributions
become important at this late stage, bringing inflation to an end
and stabilizing the moduli. We find that with moderate values for
$N$, one obtains both a sufficient amount of e-foldings and the
right size for the spectral index.
\end{abstract}

\noindent
PACS: 04.65.+e, 11.25.Mj, 11.25.Yb\\
Keywords: Inflation, M-Theory\\
hep-th/0501130

\newpage
\pagenumbering{arabic}

\section{Introduction and Summary}

Based on the recent progress of constructing de Sitter vacua in
string-theory \cite{dS1}, \cite{dS2} and M-theory \cite{BCK},
\cite{dS3} there has been an extensive effort to derive cosmic
inflation from string- and M-theory \cite{Infl}\footnote{The new
recent proposal of obtaining inflation and standard cosmology from
ghost dynamics \cite{GD} has so far not been derived from
string-theory but seems promising as the ghost has to be an axion
\cite{ACLM}.}. Most of these approaches aim to derive a
sufficiently flat potential to realize new inflation
\cite{NewInfl}. In these studies it became clear that in most
cases (if not all) a considerable degree of fine-tuning has to be
applied to achieve this goal. For a status report on the progress
see \cite{Burgess}. Closely related to the problem of obtaining a
single extremely flat slow roll direction, there is a second
problem -- the one of moduli stabilization. In order to have the
universe rolling along an extremely flat direction in moduli space
one has to ensure that the potential in all remaining directions
is sufficiently curved upwards. In other words, all moduli except
the modulus which serves as the inflaton have to be stabilized
before the inflationary phase.

\begin{figure}[t]
\begin{picture}(300,145)(0,0)


\DashCurve{(77,80)(80,60)(84,35)(86,33)(88,34)(91,40)(95,65)(98,90)}{1.5}
\Curve{(87,33.5)(95,31.5)(120,28)(180,25)}

\Text(75,125)[]{$U(\cM,\cM_a)$}
\Text(181,32)[]{$\cM$}
\Text(103,98)[]{$\cM_a$}


\DashCurve{(296,85)(320,88)(340,102)}{1.5}
\DashCurve{(296,75)(320,88)(340,95)}{1.5}
\DashCurve{(296,103)(320,88)(340,120)}{1.5}
\Curve{(320,88)(322,75)(325,60)(329,45)(334,30)(340,17)(347,7)}

\Text(305,135)[]{$U(Y_{ji},\cM_b)$}
\Text(355,14)[]{$Y_{ji}$}
\Text(354,101)[]{$\cM_b$}

\end{picture}
\caption{\it \label{Fig1} The realization of new inflation (left
figure) requires an extremely flat potential in the direction of
the modulus $\cM$ playing the role of the inflaton. This ensures
that the $\cM$ kinetic term is small and leads approximately to an
exponential expansion $a(\mathsf{t})\sim e^{H\mathsf{t}}$ where
$H\simeq \sqrt{U/3M_{Pl}^2}$. Necessarily the potential in all
other moduli directions $\cM_a$ has to be strongly curved upwards.
In contrast, for the realization of assisted inflation (right
figure) we use identical steeply decreasing exponential potentials
for many moduli $Y_{ji}$ which serve as inflatons. The outcome is
a power-law inflation $a(\mathsf{t})\sim \mathsf{t}^{p(N)}$. Since
the potential in the $Y_{ji}$ directions are the steepest
directions available during inflation, the universe will follow
their path even if the potential has a mild runaway in some of the
remaining moduli directions $\cM_b$.}
\end{figure}
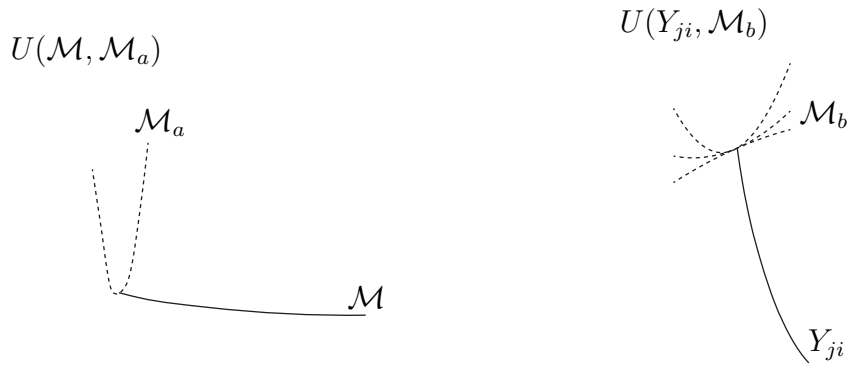

It is the goal of this paper to show that inflation can be
naturally realized in M-theory. The key for this will come from
the dynamics of many M5-branes\footnote{The cosmological
implication of a single M5-brane has been studied e.g.~in
\cite{M5}.}. Their mutual interactions, stemming from open
membrane instantons, lead to exponential potentials. These
potentials are too steep to give rise to inflation individually.
However, when taken together, they increase the Hubble friction
and lead to a specific type of inflation, known as assisted
inflation \cite{AssInfl}. It is therefore essential to have many
M5-branes present and not just one. The great advantage of
realizing inflation in M-theory in this way, is that there is no
need to stabilize all moduli before the inflationary phase. This
comes from the fact that the inflatons, which correspond to the
distances between the M5-branes in the $\Orb$ direction, turn out
to correspond to the steepest possible directions of the potential
(see fig.\ref{Fig1}). This is in striking contrast to a
realization of inflation as new inflation. The assisted type of
inflation, which is a generalization of power-law inflation
\cite{PLInfl}, has only one parameter $p$ which will be a function
of $N$, the number of M5-branes being present. As we will see,
both the number of e-foldings and the spectral index, which
characterizes the power law spectrum of primordial curvature
perturbations, lie in the right regime if we assume values around
$N \simeq 89$. Moreover, we will find a natural exit from
inflation and a reheating mechanism. The distances between the
M5-branes, which play the role of the inflatons, grow during the
inflationary phase. Once they have grown to a size comparable to
the orbifold size itself, other flux and non-perturbative
contributions to the potential become important and terminate
inflation. These other contributions, most notably gaugino
condensation on the hidden boundary and boundary-boundary open
membrane instantons, will stabilize the moduli through mechanisms
described in \cite{BCK}. Reheating occurs when the M5-branes
collide with the boundaries. In particular the M5-branes colliding
with the visible boundary will reheat our universe when they
dissolve with the boundary via small instanton transitions
\cite{SITW}, \cite{SITO}.

Let us motivate and provide some background for our approach. To
break supersymmetry spontaneously one can either use fluxes or
non-perturbative effects. The latter lead to exponential
potentials. In the context of inflation, it has been known for a
long time that simple exponential potentials of the form
\beqa
U(\varphi)
= U_0 e^{-\sqrt{\frac{2}{p}}\frac{\varphi}{M_{Pl}}} \; ,
\label{PL1}
\eeqa
with a parameter $p>1$, lead to power-law inflation \cite{PLInfl}
(by $M_{Pl}$ we denote the reduced Planck-scale). In these models
the scale-factor of the four-dimensional
Friedmann-Robertson-Walker (FRW) universe grows with cosmic time
$\mathsf{t}$ like
\beqa
a(\mathsf{t}) = a_0 \mathsf{t}^p,
\label{PL2}
\eeqa
while the inflaton $\varphi$ evolves as
\beqa
\varphi(\mathsf{t}) = \sqrt{2p} M_{Pl} \ln \bigg(
\sqrt{\frac{U_0}{p(3p-1)}} \frac{\mathsf{t}}{M_{Pl}} \bigg) \; .
\label{PL3}
\eeqa
This exact solution is valid for parameters $p>1/3$. The two
slow-roll parameters
\beqa
\epsilon = \frac{M_{Pl}^2}{2}\bigg(\frac{U'}{U}\bigg)^2 \;, \qquad
\eta = M_{Pl}^2 \frac{U''}{U} \; ,
\eeqa
where a prime indicates the derivative w.r.t.~the inflaton
$\varphi$, turn out to be constant
\beqa
\epsilon = \frac{1}{p} \; , \qquad \eta = \frac{2}{p} \; .
\label{PL4}
\eeqa
To obtain inflation, it is enough to demand that $p>1$ which
guarantees that $\ddot{a}(\mathsf{t}) > 0$. If in addition both
slow-roll parameters should be sufficiently small to meet
observational constraints, we will rather have to impose that
$p\simeq 100$, as we will see later.

In this simple example the slow-roll parameters are constant and
there is thus no exit from power-law inflation. We shall see that
when embedded into string- or M-theory this presents, however, no
problem as there are additional contributions which become
relevant eventually. These will modify the simple exponential
potential and cause inflation to end. To the reader who associates
inflation mostly with new inflation, let us remark that a period
of inflation is characterized very broadly by the requirement,
\beqa
\add(\mathsf{t}) > 0 \; ,
\eeqa
which allows many more realizations than just having an extremely
flat potential as would be required for new inflation. In
particular, we will see that inflation can also be realized with
very steep directions.

In heterotic M-theory \cite{HW}, on which we want to focus
subsequently, exponential potentials arise from open membrane
instantons \cite{Becker}, \cite{Marino}, \cite{MPS}, \cite{LOPR}
which wrap genus zero holomorphic 2-cycles on the Calabi-Yau
manifold\footnote{Corrections to F-terms coming from higher genus
instantons vanish due to holomorphy. We would like to thank
E.~Witten for pointing this out to us.}. An open membrane
instanton stretching between both boundaries (i.e.~the
$\mathbf{Z}_2$ fixed planes) leads to a superpotential
\beqa
W = h e^{-T},
\label{BBOM}
\eeqa
with $T$ the K\"ahler-modulus whose real part measures the size of
the 3-cycle covered by the open membrane instanton. In the large
volume limit this will lead, apart from power-law corrections, to
an exponential potential for the real part of $T$ with parameter
$p=1$ (a factor $\sqrt{2}$ comes from the different normalizations
of the kinetic terms for a complex and a real field). With this
value of $p$ there would be no inflation. Moreover, the inflaton
$\varphi$, which we have used before to describe power-law
inflation, has canonically normalized kinetic terms which is not
the case for $T$. Upon transforming the real part of $T$ to a
canonically normalized field $\varphi_T =
M_{Pl}\sqrt{\frac{3}{2}}\ln (T+\Tbar)$, we would end up with a
double exponential instead of a simple exponential potential. This
double exponential potential is again too steep to lead to
inflation. It is therefore not possible to generate power-law
inflation from just the boundary-boundary interaction arising from
a single open membrane instanton.

There is, however, a very interesting multi-scalar extension of
the power-law inflation scenario, called assisted inflation
\cite{AssInfl}, which can give inflation even though the
individual single-field potentials cannot. This inflation scenario
is based on $N$ scalar fields $\varphi_i,\; i=1,\hdots,N$, each of
which possesses a potential
\beqa
U = U_0 e^{-\sqrt{\frac{2}{p}}\frac{\varphi_i}{M_{Pl}}} \; ,
\qquad \forall i=1,\hdots,N \; .
\eeqa
One can map this multi-field problem to the single field power-law
problem and show that it leads as well to a power-law solution for
$a(\mathsf{t})=a_0 \mathsf{t}^{p(N)}$, where now \cite{AssInfl}
\beqa
p \rightarrow p(N) = Np \; .
\eeqa
This solution is valid if $p(N) > 1/3$ and leads to inflation for
$p(N) > 1$. Hence, even if the single exponential contributions
are too steep to support inflation individually, nevertheless one
can obtain inflation by choosing $N$ large enough. We will see
that this type of inflation arises naturally in the large volume
limit from the dynamics of $N$ M5-branes distributed along the
$\Orb$ orbifold interval.

We want to stress that there is one very important benefit when
inflation is realized in string- or M-theory through assisted
inflation. This benefit concerns the issue of moduli
stabilization. We will see that the canonically normalized
inflaton fields $\varphi_i$ originate from the real parts of
moduli $Y_{ji}=Y_j-Y_i,\; i,j =1,\hdots,N$ which describe the
position difference of the $N$ M5-branes along the $\Orb$
interval. It is the $Y_{ji}$ directions in which the potential
decreases fastest. During the inflationary period these are the
only directions where the potential falls off exponentially fast,
$U\sim e^{-Y_{ji}-\Ybar_{ji}}$. Hence, this alleviates
considerably the task to have all other moduli, except the
inflatons, stabilized before the inflationary phase. We are no
longer dealing with an extremely flat direction but instead with
the opposite extreme, the steepest possible directions (see
fig.\ref{Fig1}). To ensure that the universe is indeed following
this steepest path, it is enough to require that all other
directions are less steep than this exponential decrease. This is,
however, a condition which is much easier met and could also allow
for a mild runaway in the non-inflating directions.

The organization of this paper is as follows. In section 2 we
explain the heterotic M-theory setup and derive the potential with
$N$ parallel M5-branes being present. We argue that during the
inflationary phase the dominant contribution to the potential
originates from the forces between the M5-branes. These arise from
open membrane instantons stretching between pairs of M5-branes. In
section 3 we describe the regime in moduli space where we can map
the M5-brane dynamics to the dynamics of assisted inflation.
During inflation the distances between the M5-branes, which
represent the inflatons, will grow until the M5-branes coalesce
with the boundaries via small instanton transitions. This will
partly reheat the universe. Towards the end of inflation the
orbifold size will start to grow, pushing the hidden gauge theory
to strong coupling. The onset of gaugino condensation on the
hidden boundary and the ensuing non-trivial $H$-flux will
stabilize in particular the orbifold size. These additional
contributions to the potential will terminate inflation and the
implied stabilization of the orbifold modulus will contribute to
the reheating, as well. In section 4 we describe in more detail
the exit from inflation and derive the implications of our
M-theory inflation proposal for the spectral index and the number
of e-foldings. With a moderate number of M5-branes both of these
quantities comply with their observationally derived values.

\section{The Multi M5-Brane Potential}

We are focussing in this work on embedding inflation into
heterotic M-theory. As motivated in the introduction, our aim will
be to show that assisted inflation can arise in M-theory. To this
end we have to find a setup with several scalar fields, having
each the same exponential potential. Let us therefore consider
heterotic M-theory in the presence of $N$ parallel M5-branes
distributed along the $\Orb$ orbifold interval. When compactified
down to four dimensions on a six-manifold preserving N=1
supersymmetry in four dimensions, the background is given through
a warping of the six-manifold along the $\Orb$ interval
\cite{WWarp}, \cite{CK1}. In these solutions the six-manifold can
either be a Calabi-Yau manifold or a more general non-K\"ahler
manifold \cite{Strom}, \cite{Beckernk}, \cite{CL}, the relevant
warping for both cases will be the same. For simplicity, we will
take a Calabi-Yau manifold henceforth. All $N$ M5-branes will fill
the four-dimensional non-compact spacetime and wrap the same
holomorphic two-cycle $\Sigma_2$ on the Calabi-Yau. To illustrate
the mechanism in its simplest form, we will work with one K\"ahler
modulus $T$ only, hence take $h^{1,1}=1$.

The effective four-dimensional N=1 supergravity theory is
described in terms of the volume modulus $S$ of the Calabi-Yau,
the modulus associated to the length of the $\Orb$ orbifold $T$
and the M5-brane position chiral superfields $Y_i$ (we are
suppressing the gauge bundle moduli related to the $E_8$
Yang-Mills gauge sectors)
\begin{alignat}{3}
S &= \cV + \cV_{OM}\sum_{i=1}^N
\Big(\frac{x_i^{11}}{L}\Big)^2 + i \sigma_S \;, \\
T &= \cV_{OM} + i\sigma_T \;, \\
Y_i &= \cV_{OM} \Big(\frac{x_i^{11}}{L}\Big) + i \sigma_i \;,
\qquad i=1,\hdots,N \; .
\end{alignat}
Here $\cV$ denotes the Calabi-Yau volume averaged over $\Orb$ and
$\cV_{OM}$ the averaged volume of an open membrane instanton
wrapping $\Sigma_2$ and stretching from one boundary to the other.
$L$ is the length of the $\Orb$ interval and the position modulus
of the $i$th M5-brane ranges over $0\le x_i^{11}\le L$. The axions
$\sigma_S, \sigma_T, \sigma_i$ arise from various components of
the three-form potential $C$ of eleven-dimensional supergravity,
see \cite{CK2}. Note the correction to the real part of $S$ which
arises through the presence of the M5-branes \cite{DS},
\cite{MPS}. A similar correction arises in the K\"ahler-potential.
A priori each M5-brane could wrap $\beta_i$ times the basis
two-cycle $\Sigma_2$. The above expressions assume, for
simplicity, that all $\beta_i$ are equal to one. Furthermore,
there are $h^{2,1}$ complex structure moduli $Z^\alpha$. Let us
define in addition the real moduli\footnote{Note the typographic
difference between cosmic time $\mathsf{t}$ and the modulus $t$.}
\begin{alignat}{3}
s = S+\Sbar, \qquad t &= T+\Tbar, \qquad
y_i = Y_i+\Ybar_i, \\
y &= \Big( \sum_{i=1}^N y_i^2 \Big)^{1/2} \;.
\end{alignat}
It will be convenient to further define
\begin{alignat}{3}
Q &= s-\frac{y^2}{t}, \\
R &= 3Q^2-2\frac{y^4}{t^2} \; ,
\end{alignat}
which will appear in the K\"ahler-potential and the matrix of its
second derivatives. We will see shortly that in order to have a
well-defined K\"ahler potential and a positive definite K\"ahler
metric $K_{I\Jbar}$ of second derivatives, both $Q$ and $R$ have
to be positive. The requirement of having $R>0$ is more stringent
and imposes the following restriction on the $s,t,y$ moduli
\beqa
(3-\sqrt{6}) s > \frac{y^2}{t} \; .
\label{Cnt1}
\eeqa

The K\"ahler-potential for these moduli \beqa K =
K_{(S)}+K_{(T)}+K_{(Y)}+K_{(Z)}, \eeqa is given by \cite{LOW1},
\cite{DS}, \cite{MPS}
\begin{alignat}{3}
K_{(S)} + K_{(Y)} &= -\ln \Big(S+\Sbar - \frac{\sum_{i=1}^N
(Y_i+\Ybar_i)^2}{T+\Tbar}\Big), \\
K_{(T)} &= -\ln\Big(\frac{d}{6}(T+\Tbar)^3\Big), \\
K_{(Z)} &= -\ln\Big(i\int_{CY} \Om\we\Ombar \Big), \;
\end{alignat}
with $d$ the Calabi-Yau intersection number. Since, we can more
succinctly write
\beqa
K_{(S)} + K_{(Y)} = -\ln Q \; ,
\eeqa
it is indeed clear that $Q$ has to be positive \beqa Q > 0 \; .
\eeqa
Another way to see this is to rewrite $Q=2\cV$ and to note that in
the backgrounds of \cite{CK1} the average Calabi-Yau volume is
always positive, as it should. Moreover, one finds that the
determinant of the ensuing K\"ahler metric is (the indices
$I,J,\hdots$ run over all complex moduli)
\beqa
\det K_{I\Jbar} = \frac{16 R}{Q^{2N} t^6} \det G_{\alpha\betabar}
\; ,
\eeqa
with $G_{\alpha\betabar}$ the metric on the complex
structure $Z^\alpha$ moduli space. Therefore, also $R$ has to be
positive
\beqa
R > 0 \; ,
\eeqa
to ensure a positive definite K\"ahler metric $K_{I\Jbar}$.

Let us next discuss the relevant superpotential. During the epoch
of inflation we will assume vanishing vacuum expectation values
for charged matter fields. The corresponding trilinear Yukawa
superpotential and higher-order perturbative boundary-boundary
contributions \cite{K1} will therefore be absent. The remaining
contributions to the superpotential will come from open membrane
instantons wrapping each the same $\Sigma_2$ on the Calabi-Yau,
and stretching either between both boundaries (99), between two of
the M5-branes (55), between the visible boundary and an M5-brane
(95) or between an M5-brane and the hidden boundary (59)
\beqa
W_{OM}
= W_{99}
+ W_{55}
+ W_{95} + W_{59} \; .
\eeqa
These superpotentials are given by
\begin{alignat}{3}
W_{99} &= h e^{-T}, \\
W_{95} &= h \sum_{i=1}^N e^{-Y_i}, \\
W_{59} &= h \sum_{i=1}^N e^{-(T - Y_i)}, \\
W_{55} &= h \sum_{i<j} e^{-Y_{ji}} \; ,
\end{alignat}
where
\beqa
Y_{ji} = Y_j - Y_i \; ,
\eeqa
describe the difference in location of the $j$th and the $i$th
M5-brane. Usually, one would also consider gaugino condensation
\cite{Dine} on the hidden boundary \cite{LOW4}
\beqa
W_{GC} = -C_H \mu^3 e^{-\frac{1}{C_H}f_h} \; ,
\qquad f_h = S+\gamma_h T+\frac{\sum_i \gamma_i Y_i^2}{T},
\eeqa
where $C_H$ is the dual Coxeter number related to the hidden gauge
group and $\mu$ is determined in terms of the ultraviolet cut-off
scale for the hidden gauge theory, see \cite{BCK}. The parameters
$\gamma_h,\gamma_i$ are defined as
\beqa
\gamma_{h,i} = \beta_{h,i} \frac{\pi L}{V_v}
\Big(\frac{\kappa}{4\pi}\Big)^{2/3} \int_{\Sigma_2}\om \; ,
\eeqa
where $\om$ is the K\"ahler form of the Calabi-Yau and ${V_v}$ the
Calabi-Yau volume at the location of the visible boundary. All
$\beta_i=1$, as explained before, and the integers $\beta_h$ are
obtained as expansion coefficients of the second Chern classes of
the hidden boundary vector $F_h$ and the Calabi-Yau's tangent
bundle $TX$
\beqa
c_2(F_h)-\frac{1}{2}c_2(TX) = \beta_h [\Sigma_2] \; .
\eeqa
Similarly, one obtains the integer $\beta_v$ from the
corresponding visible boundary bundles. The anomaly cancelation
equation demands
\beqa
\beta_v+\beta_h+\sum_{i=1}^N\beta_i =
\beta_v+\beta_h+N = 0 \; .
\eeqa
Via a perfect square structure of the M-theory action, gaugino
condensation implies a non-vanishing NS-NS three-form field-strength
$H$ of type $(3,0)+(0,3)$ on the hidden boundary \cite{HGC}. Hence,
once gaugino condensation sets in, a perturbative flux superpotential
\beqa W_H = \int_{CY_h} H \we \Om, \eeqa integrated over the
Calabi-Yau on the hidden boundary has to be taken into account as
well.

\begin{figure}[t]
\begin{picture}(300,145)(0,0)

\LongArrow(210,0)(50,0)
\LongArrow(240,0)(395,0)
\Text(225,0)[]{$L$}

\Text(45,125)[]{$E_8$}
\Text(405,125)[]{$E_8$}
\Text(230,100)[]{M5's}


\Line(40,-5)(40,105)
\Line(50,5)(50,115)
\Line(40,-5)(50,5)
\Line(40,105)(50,115)

\Line(400,-5)(400,105)
\Line(410,5)(410,115)
\Line(400,-5)(410,5)
\Line(400,105)(410,115)


\Line(200,15)(200,80)
\Line(210,85)(210,90)
\Line(200,15)(205,20)
\Line(200,80)(210,90)

\Line(205,15)(205,80)
\Line(215,85)(215,90)
\Line(205,15)(210,20)
\Line(205,80)(215,90)

\Line(210,15)(210,80)
\Line(220,85)(220,90)
\Line(210,15)(215,20)
\Line(210,80)(220,90)

\Line(215,15)(215,80)
\Line(225,85)(225,90)
\Line(215,15)(220,20)
\Line(215,80)(225,90)

\Line(220,15)(220,80)
\Line(230,85)(230,90)
\Line(220,15)(225,20)
\Line(220,80)(230,90)

\Line(225,15)(225,80)
\Line(235,85)(235,90)
\Line(225,15)(230,20)
\Line(225,80)(235,90)

\Line(230,15)(230,80)
\Line(240,85)(240,90)
\Line(230,15)(235,20)
\Line(230,80)(240,90)

\Line(235,15)(235,80)
\Line(245,85)(245,90)
\Line(235,15)(240,20)
\Line(235,80)(245,90)

\Line(240,15)(240,80)
\Line(250,85)(250,90)
\Line(240,15)(245,20)
\Line(240,80)(250,90)

\Line(245,15)(245,80)
\Line(255,25)(255,90)
\Line(245,15)(255,25)
\Line(245,80)(255,90)

\end{picture}
\caption{\it\label{Fig2} At the beginning of the inflationary
epoch, we assume all $N$ M5-branes to be grouped around some
common location on the $\Orb$ interval such that the open membrane
instanton interactions between the M5-branes dominate the
potential.}
\end{figure}
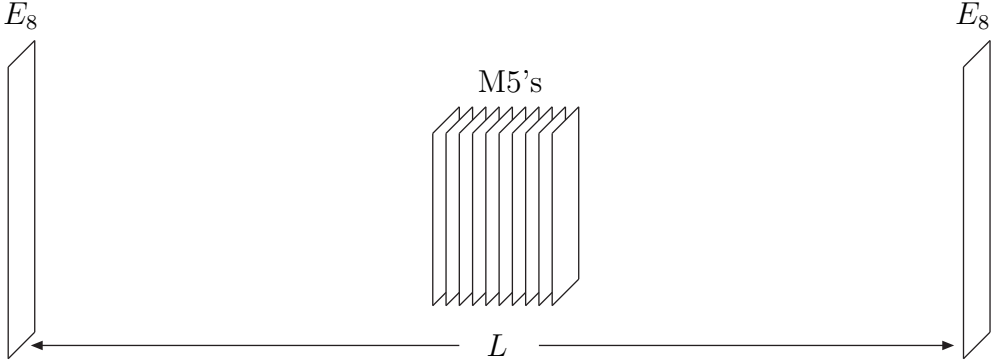

It is important to recall under which geometrical conditions
gaugino condensation on the hidden boundary has to be taken into
account and when it should be omitted. The background geometry of
heterotic M-theory \cite{CK1} is warped along the $\Orb$
exhibiting a singularity at some finite coordinate distance where
the warp-factor vanishes. Phenomenological considerations
\cite{WWarp}, \cite{CK1} make it desirable to place the hidden
boundary right at this singularity, which implies a strongly
coupled hidden gauge theory in which gaugino condensation sets in.
Actually, the hidden boundary needs not be placed at this critical
distance by hand, but can indeed be stabilized here through
stabilization of the orbifold length modulus $T$ \cite{CK2},
\cite{BCK}. This situation describes successfully the particle
phenomenology as we witness it today \cite{HetMPheno1},
\cite{HetMPheno2} leading at the same time to dark matter
candidates from hidden matter \cite{KSUSY03}. However, the
situation in the early universe during the epoch of inflation
might be different. In particular, it is conceivable that the
orbifold size starts off at a subcritical
value\footnote{Subcritical means smaller than the critical
orbifold size which would place the hidden boundary right on top
of the singularity.} and grows only gradually to its critical
value towards the end of inflation. Indeed to achieve a
stabilization at the critical orbifold size, the hidden $E_8$
gauge group must already be broken down to a group with
considerably smaller $C_H$ \cite{BCK}, \cite{KSUSY03}. Therefore,
if at the beginning of inflation, we assume an unbroken $E_8$ on
the hidden boundary, we could stabilize the orbifold size only at
subcritical value if the hidden gauge theory would remain strongly
coupled. One has to notice, however, that when bringing the hidden
boundary to subcritical distance, the hidden gauge theory will
soon become perturbative since the Calabi-Yau volume on the hidden
boundary quickly grows when the orbifold size shrinks \cite{CK1}.
More quantitatively, with a visible gauge coupling
$\alpha_v=1/25$, the hidden gauge coupling $\alpha_h$ will be
smaller than $1$ (smaller than $1/2$), if the orbifold size is
less than $0.80$ (less than $0.72$) of its critical size. Under
the assumption that at the beginning of inflation we start with
such a subcritical orbifold size (we will see that this condition
remains upright during inflation), it is therefore no longer
justified to take gaugino condensation into account and we will
consequently omit it. Necessarily, we then have to omit during
inflation the $H$-flux superpotential induced by it, as well. Both
contributions will, however, become important when the orbifold
size grows towards the end of inflation to its critical size. We
will therefore only take $W_{OM}$ as our superpotential. Among
these open membrane interactions we want to focus on those between
the M5-branes, i.e.~take for the superpotential
\beqa
W = W_{55} \; .
\eeqa
This can easily be achieved by having initially all $N$ M5-branes
grouped together around some common generic site, not too close to
either boundary, along the $\Orb$ interval (see fig.\ref{Fig2}).
Then the inter-boundary interaction $W_{99}$ and the interactions
between M5-branes and either boundary $W_{95}, W_{59}$ can be
neglected since the corresponding open membrane instantons have to
stretch over longer distances.

The potential can then be calculated from the N=1 supergravity
expression for F-terms contributions
\beqa
U = M_{Pl}^4 e^{K}\left(\sum K^{\Ibar J}D_{\Ibar}\Wbar_{55} D_J
W_{55} - 3 |W_{55}|^2 \right) \; ,
\eeqa
which gives ($D_i W_{55} \equiv \partial W_{55}/\partial Y_i +
W_{55}\partial K/\partial Y_i$)
\begin{alignat}{3}
\frac{U}{M_{Pl}^4 e^K} = \; &G^{\alphabar\beta}D_{\alphabar}W_{55}
D_\beta W_{55}+ Qt\sum_{i,j=1}^N
\Big(\frac{1}{2}\delta_{ij}+\frac{Q}{Rt}y_iy_j\Big)
\overline{D_i W_{55}} D_j W_{55} \notag \\
&+\Big( \frac{3Q^2}{R} -\frac{2y^2}{Qt} \Big) |W_{55}|^2 \; .
\label{FullPot}
\end{alignat}
The multiplying K\"ahler factor is
\beqa
e^K =
\frac{6}{(i\int \Om\we\Ombar)Qt^3d} \; .
\eeqa
Let us note that the second line of the potential comes from the
terms
\beqa
\sum K^{\Ibar J} K_{\Ibar} K_J |W_{55}|^2 \; ,
\eeqa
with the sum running over all moduli, combined with the
$-3|W_{55}|^2$ term. One can check that $3Q^2/R > 2y^2/Qt$,
i.e.~the second line of the potential gives a positive
contribution, if $y < 1.83\sqrt{st}$. Otherwise its contribution
is negative. Now the moduli constraint (\ref{Cnt1}) implies that
$y < \sqrt{3-\sqrt{6}}\sqrt{st}=0.74\sqrt{st}$. Hence, the second
line and therefore the whole potential will be {\em positive}.
This is an important requisite for a derivation of power-law
respectively assisted inflation.

\section{Inflation from Multi M5-Brane Dynamics}

\subsection{The Inflationary Regime}

We will now specify more precisely the regime in moduli space
where inflation occurs, i.e.~where we can map the dynamics of the
interactions between the M5-branes to the dynamics leading to
assisted inflation. First of all, we will impose the constraints
\begin{alignat}{3}
D_\alpha W_{55} &= 0 \; , \label{CovDer1} \\
D_i W_{55} &= 0 \label{CovDer2} \; .
\end{alignat}
Since we have just found that the potential is positive definite
in these derivatives, abandoning these constraints would increase
the potential energy and is thus disfavored on dynamical grounds.
These two constraints guarantee therefore a partial minimization
of the potential. Let us now study their implications. The first
constraint is equivalent to
\beqa
\frac{\partial \ln h}{\partial Z^\alpha} =  -\frac{\partial
K_{(Z)}}{\partial Z^\alpha} \; .
\label{Cnt2}
\eeqa
This implies
\beqa
h = i\int_{CY} \Omega\we\Ombar \; ,
\label{Fix1}
\eeqa
up to a $Z^\alpha$ independent integration constant which we have
set to zero. The $h^{2,1}$ conditions (\ref{Cnt2}) will fix the
complex structure moduli.

Before discussing the implication of the second constraint, let us
first derive an important upper bound on $N$. After applying the
vanishing of the two K\"ahler-covariant derivatives, the potential
$U$ reads
\beqa
\frac{d(i\int \Om\we\Ombar)}{6M_{Pl}^4} U = \Big( \frac{3Q}{Rt^3}
-\frac{2y^2}{Q^2t^4} \Big) |W_{55}|^2 \; .
\label{Pot2}
\eeqa
If this potential should be mapped to an assisted inflation
dynamics with the inflatons arising from the $y_i$ differences,
then we have to make sure that the inflatons appear only in the
exponentials of $W_{55}$. In other words, during inflation the
prefactor in brackets shouldn't depend on the inflatons and
therefore not on $y$. This requirement can be met when the
condition
\beqa
Qt \gg y^2 \; ,
\label{Lim1}
\eeqa
is satisfied. It implies
\begin{alignat}{3}
&Q \simeq s \label{Imp2} \; ,\\
&R \simeq 3Q^2 \simeq 3s^2 \label{Imp3} \; ,
\end{alignat}
and also that $3Q/Rt^3 \gg 2y^2/Q^2t^4$. Hence one can neglect the
second term in the potential's prefactor and the potential becomes
\beqa
\frac{d(i\int \Om\we\Ombar)}{6M_{Pl}^4} U = \frac{1}{st^3}
 |W_{55}|^2 \; ,
\label{Pot3}
\eeqa
with a $y$ independent prefactor as desired.

It is interesting that the condition (\ref{Lim1}) amounts to an
upper bound on $N$, the number of M5-branes. This narrows the
range for $N$ considerably, as we will see, and increases the
mechanism's predictability. Indeed, this bound is tighter than the
bound (\ref{Cnt1}) for a consistent supergravity, which becomes
$(3-\sqrt{6})Qt>y^2$. Let us add that on the other hand, we also
know that with a too small $N$ we won't be able to derive an
assisted inflation mechanism. Hence the range for $N$ giving
inflation, is limited from above and also non-trivially from
below.

After this preparation we can now address the geometrical meaning
of the vanishing of the K\"ahler covariant derivative
\beqa
D_i W_{55} = W_{55,i} + \frac{2y_i}{Qt}W_{55} = 0 \; .
\label{CovDer3}
\eeqa
To trust the effective supergravity description, we have to work,
as always, in the regime where $s,t$ are both considerably larger
than one. We then find that $Qt\simeq st \gg t > y_i$. Hence, we
can neglect the second term in the covariant derivative, reducing
it to an ordinary partial derivative, $D_i W_{55}
\rightarrow W_{55,i}$. To see most clearly what the vanishing of
this derivative implies, let us concentrate from now on on the
dominant nearest neighbor interactions between adjacent M5-branes,
i.e.~set
\beqa
W_{55} = h \sum_{i=1}^{N-1} e^{-Y_{i+1,i}} \; .
\label{NNInt}
\eeqa
The dominance of the nearest neighbor interactions is valid if the
M5-branes are roughly equidistantly distributed. An exact
equidistant distribution is however precisely what the vanishing
of the derivative $W_{55,i}$, and therefore energy minimization,
implies
\beqa
Y_{i+1,i} \equiv \Delta Y \; .
\label{ComDiff}
\eeqa
All complex nearest-neighbor differences have the same value (see
fig.\ref{Fig2}) and will be set equal to a common $\Delta Y$
henceforth.

The equidistant M5-brane distribution, which makes the
nearest-neighbor M5-brane interactions the dominant ones, allows
us to derive for the potential $U$ the following simple expression
\beqa
\frac{U d}{6M_{Pl}^4 (i\int \Om\we\Ombar)} = \frac{(N-1)^2}{s t^3}
e^{-\Delta y} \; .
\label{Pot4}
\eeqa
Here we have defined the real distance modulus
\beqa
\Delta y = \Delta Y + \overline{\Delta Y} \; ,
\eeqa
and employed the relation (\ref{Fix1}) to eliminate $h$.

\subsection{Mapping the M5-Brane Dynamics to Assisted Inflation}

Finally we have to transform the M5-brane position fields $Y_i$ to
fields whose kinetic terms are canonically normalized. The reason
is that the assisted inflation dynamics is formulated in terms of
canonically normalized scalar fields. The kinetic term for the
$Y_i$ is
\beqa
S_{kin} = -M_{Pl}^2 \int d^4x \sqrt{-g} K_{i\jbar}
\partial_\mu Y_i \partial^\mu \Ybar_{\jbar} \; ,
\eeqa
with a summation over indices understood, where
\beqa
K_{i\jbar} = \frac{4y_iy_j+2Qt\delta_{ij}}{Q^2t^2} \; .
\eeqa
It follows from (\ref{Lim1}) that $Qt\gg y^2 = \sum y_i^2 > y_i
y_j$. Hence the first term in the numerator becomes negligible and
we get $K_{i\jbar}=2\delta_{ij}/Qt$. The canonically normalized
complex fields are therefore $M_{Pl}\sqrt{2/Qt} Y_i$ leading to
the canonically normalized real M5-brane position and difference
fields
\beqa
\phi_i =
\frac{2M_{Pl}}{\sqrt{Qt}} y_i ,
\qquad \Delta \phi = \frac{2M_{Pl}}{\sqrt{Qt}} \Delta y \; .
\eeqa

We have just seen that a potential is only generated for the
distance modulus between nearest neighbor M5-branes but not for
their combined center-of-mass position. Let us therefore now
switch from the $N$ position fields $\phi_i$ to the more adequate
description in terms of the M5-brane center-of-mass field
\beqa
\phi_{cm} = \frac{1}{N}(\phi_1 + \hdots + \phi_N) \; ,
\eeqa
and the difference field $\Delta\phi$. The relation between the
two sets of fields is provided by the relation
\beqa
\phi_i = \phi_{cm} + \Big(i-\frac{N+1}{2}\Big)\Delta\phi \; ,
\eeqa
which is derived in the appendix. Since there is no potential for
$\phi_{cm}$, its value will stay constant and its kinetic term
vanishes. The sum of the $\phi_i$ kinetic terms thus becomes
\begin{alignat}{3}
\frac{1}{2}\sum_{i=1}^N \partial_\mu\phi_i\partial^\mu\phi_i &=
\partial_\mu\Delta\phi\partial^\mu\Delta\phi \sum_{i=1}^N
\Big(i-\frac{N+1}{2}\Big)^2 \\
&= \frac{N(N^2-1)}{12}
\partial_\mu\Delta\phi\partial^\mu\Delta\phi \; ,
\end{alignat}
which requires us to perform a second rescaling. The canonically
normalized difference field, denoted now by $\varphi$, with
standard kinetic term $(1/2)\partial_\mu \varphi \partial^\mu
\varphi$ becomes
\beqa
\varphi = \sqrt{\frac{N(N^2-1)}{6}}\Delta\phi
= M_{Pl} \sqrt{\frac{2N(N^2-1)}{3Qt}}\Delta y \; .
\eeqa

We are now ready to map the dynamics of the single remaining
difference field $\varphi$ to the power-law inflation dynamics
given in the introduction. For this, let us notice that written in
terms of $\varphi$, we have found a potential (\ref{Pot4})
\beqa
U (\varphi) = \tilde{U}_0 (N-1)^2
e^{-\sqrt{\frac{3Qt}{2N(N^2-1)}}\frac{\varphi}{M_{Pl}}},
\label{Pot4}
\eeqa
where
\beqa
\tilde{U}_0 = \frac{6M_{Pl}^4 (i\int\Om\we\Ombar)}{s t^3 d} \; .
\eeqa
This positive valued prefactor can be regarded as being
approximately constant throughout the inflationary period. The
reason is that there are no steep potentials being present for
$s,t$ during the inflationary period where (\ref{Pot4}) is valid.
The potential for $t$ arising from $W_{99}$ is strongly suppressed
against the M5-brane potential arising from $W_{55}$ and gaugino
condensation. Gaugino condensation would deliver a steep potential
for $s$ but is absent as argued before. Therefore the size of the
orbifold will stay approximately constant during the inflationary
M5-brane evolution which we consider here.

In a spatially flat four-dimensional FRW universe we then have a
Hubble parameter
\beqa
H^2 = \frac{1}{3M_{Pl}^2}\Big(U(\varphi) +
\frac{1}{2}\dot{\varphi}^2\Big) \;,
\eeqa
and the dynamics of $\varphi$ is determined through
\beqa
\ddot{\varphi}+3H\dot{\varphi}+\frac{dU}{d\varphi} = 0 \; .
\eeqa
This is precisely the dynamics which leads to power-law inflation
\cite{PLInfl} (see also \cite{LLBook}). To achieve the mapping to
power-law inflation, (\ref{PL1}), we merely need to set
\begin{alignat}{3}
p &= \frac{4N(N^2-1)}{3Qt} \; , \label{PLIp} \\
U_0 &= \tilde{U}_0 (N-1)^2 \;.
\end{alignat}
Hence, we end up with a power-law evolution for the scale-factor
(\ref{PL2}) and the solution (\ref{PL3}) for $\varphi$. As
discussed earlier, we need $p > 1$ to obtain power-law inflation.
Hence, we have to impose the further constraint
\beqa
p > 1 \qquad \Leftrightarrow
\qquad 4N(N^2-1) > 3Qt \; .
\label{Lim2}
\eeqa

During the inflationary phase both $Q$ and $t$ stay approximately
constant and $y$ does not vary much since the M5-branes spread
towards both boundaries. Therefore the two conditions (\ref{Lim1})
and (\ref{Lim2}) remain valid during inflation, if they have been
fulfilled initially. Moreover, since during inflation the M5-brane
nearest-neighbor distances grow in an equidistant way, also
(\ref{CovDer2}) remains satisfied. The same is of course true for
(\ref{CovDer1}). This is no surprise as we have seen that the
vanishing of both K\"ahler covariant derivatives minimizes the
energy and are thus dynamically selected. We can therefore
conclude that all the conditions which are needed to generate
inflation, do not break down during the inflationary process.

This is a very interesting result, as it shows that assisted
inflation can be realized successfully in heterotic M-theory. The
embedding of assisted inflation into string theory was explored
previously in the type IIB context in \cite{FM}. There it was
found that tree level potentials resulting from fluxes do not
induce potentials that lead to inflation and it was speculated
that instanton corrections may lead to such a potential. This is
precisely what we have shown herein. Indeed type IIB brane-world
models proposed in \cite{K3} have a two boundary setup very
similar to the $\Orb$ setup studied here. With the role of the
M5-branes played by D5 or D7-branes it might thus be possible to
transfer the M-theory inflation mechanism also to type IIB within
this brane-world framework.

\section{Exit from Inflation and Observational Results}

So far we have demonstrated that inflation in M-theory can arise
from multi M5-brane dynamics. The idea is to have initially all
M5-branes rather close together such that their dynamics dominates
the potential. In particular the boundary-boundary interactions
$W_{99}$ are negligible and the growth of the distances between
adjacent M5-branes will be much more rapid than the growth of any
other modulus. The ensuing dynamics can be mapped to power-law
inflation, leading to a sustained period of inflation in the
regime determined by the two constraints (\ref{Lim2}),
(\ref{Lim1})
\beqa
\frac{4}{3}N(N^2-1) > Qt \gg y^2 \; ,
\label{Lim}
\eeqa
The left inequality ensures $p > 1$ and therefore guarantees
inflation, whereas the right inequality led to the simple
exponential potential (\ref{Pot4}) required for power-law
inflation. From these constraints it is clear, that we wouldn't
have obtained inflation if just one or too few M5-branes would
have been present. In that case $p$ would be much smaller than 1
and cannot give inflation. To obtain power-law inflation from a
multitude of equal exponential potentials which are too steep to
give inflation on their own, but will do so when considered
together due to an increased Hubble friction, is the idea of
assisted inflation \cite{AssInfl}. It is thus an assisted
inflation mechanism which we have derived here from M-theory.

Let us comment on the naturalness of the initial configuration of
M5-branes. The positive potential (\ref{FullPot}) is in particular
positive definite in the K\"ahler covariant derivatives.
Therefore, setting $D_i W_{55}=0$ already minimized partially the
energy and is therefore dynamically motivated. In the large volume
regime, in which we are working, this condition became simply
$W_{55,i}=0$. When considering just nearest-neighbor open membrane
interactions, the geometrical meaning of this equation was
precisely that the M5-branes had to be equidistantly distributed.
Therefore, among the many initial M5-brane distributions, the
equidistant ones are dynamically favored. It remains to find a
selection principle for the initial smallness of the common
nearest-neighbor distance and the initial localization of the
stack of M5-branes away from the boundaries. While the former
issue constitutes a fine-tuning for the time being, it can be
shown that the latter issue of having all M5-branes localized away
from the boundaries can actually be relaxed to an initial
equidistant configuration along the whole interval. This will be
dealt with in a separate publication.

In the rest of this section we would like to address the two main
observational implications and the exit from inflation. An
important quantity predicted by an inflation model is the spectral
index $n$. It determines the power-law spectrum of the primordial
curvature perturbation $\cP_{\cR}(k)\sim k^{n-1}$. From this
spectrum the spectrum of any other perturbation can be obtained by
simple multiplication with the square of the appropriate transfer
function \cite{LLBook}. For power-law inflation the spectral index
$n$ is given by
\beqa
n = 1 - \frac{2}{p} \; .
\eeqa
Observations lead to the constraint \cite{Data}\footnote{We are
grateful to R.~Kallosh for bringing this reference to our
attention.}
\beqa
n = 0.98 \pm 0.02 \; ,
\eeqa
which implies
\beqa
p \simeq 100 \; .
\label{ConObs}
\eeqa
We will now see that we can account for this observational
constraint within the regime where our derivation is valid.

To this end, let us first make both constraints obtained so far
concrete by adopting typical values $\cV = 341, \cV_{OM} = 7$
(cf.~the values in table 1 of \cite{BCK} for the relevant case of
a hidden unbroken $E_8$) and $x_i^{11}/L=\cO(1/2)$. These imply $s
= 682+3.5N, t=14, y^2\simeq 49N$. The constraints (\ref{Lim}) then
deliver the following bound on $N$
\beqa
19 < N \ll 195 \; .
\eeqa
With $N$ in this regime we will obtain inflation. Let us next
verify that the spectral index observational constraint indeed
amounts to an $N$ within this range. For this we evaluate
(\ref{PLIp}) with the same $\cV,\cV_{OM}$ as before, which gives
\beqa
p \simeq \bigg(\frac{N}{19.3}\bigg)^3 \; .
\eeqa
The spectral index constraint (\ref{ConObs}) then amounts to
\beqa
N \simeq 89 \; ,
\eeqa
which indeed lies in the above interval. Thus, without having to
invoke extremely large values for $N$, we can account for the
correct size of the spectral index $n$.

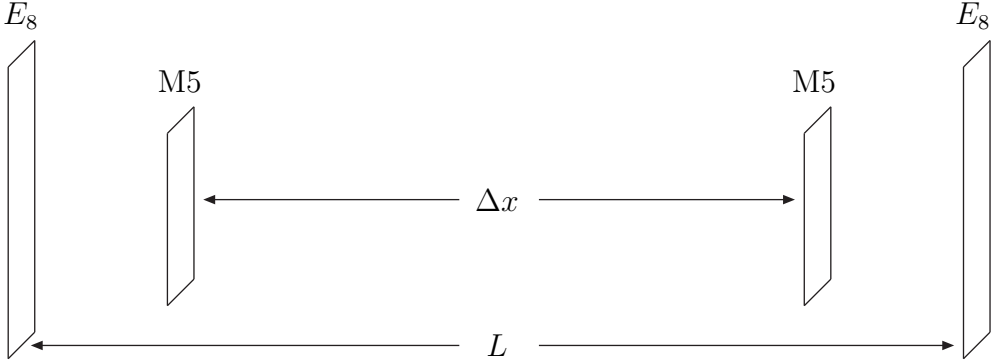
\begin{figure}[t]
\begin{picture}(300,145)(0,0)

\LongArrow(210,0)(50,0)
\LongArrow(240,0)(395,0)
\Text(225,0)[]{$L$}

\LongArrow(210,55)(115,55)
\LongArrow(240,55)(335,55)
\Text(225,55)[]{$\Delta x$}

\Text(45,125)[]{$E_8$}
\Text(405,125)[]{$E_8$}

\Text(105,100)[]{M5}
\Text(345,100)[]{M5}


\Line(40,-5)(40,105)
\Line(50,5)(50,115)
\Line(40,-5)(50,5)
\Line(40,105)(50,115)

\Line(400,-5)(400,105)
\Line(410,5)(410,115)
\Line(400,-5)(410,5)
\Line(400,105)(410,115)


\Line(100,15)(100,80)
\Line(110,25)(110,90)
\Line(100,15)(110,25)
\Line(100,80)(110,90)

\Line(340,15)(340,80)
\Line(350,25)(350,90)
\Line(340,15)(350,25)
\Line(340,80)(350,90)

\end{picture}
\caption{\it\label{Fig3} Inflation comes to an end when the
distance between adjacent M5-branes has grown to a size comparable
to the orbifold size itself. At this stage most of the M5-branes
have coalesced with the boundaries through small instanton
transitions. This reheats partly the visible boundary and
therefore our universe.}
\end{figure}

Before addressing the next important quantity, the number of
e-foldings, it will be necessary to describe first how and when
the exit from inflation occurs. The simple exponential potential
which we have found, remains valid as long as the other
contributions to the potential, gaugino condensation, $H$-flux,
and the $99,59,95$ open membrane instantons are absent or remain
negligible. In particular this requires that the hidden gauge
theory shouldn't be strongly coupled during inflation to avoid
gaugino condensation to set in. This can be easily achieved by
starting with a subcritical orbifold length at the beginning of
inflation. Moreover, we have to make sure that towards the
beginning the M5-branes have nearest neighbor distances which are
much smaller than the distance between an M5-brane to either
boundary (see fig.\ref{Fig2}). This ensures that it is indeed the
$55$ M5-brane sector which dominates the potential. What happens
during the inflationary phase is that the coordinate distance
\beqa
\Delta x(\mathsf{t}) \equiv x_{i+1}^{11}(\mathsf{t}) -
x_i^{11}(\mathsf{t})
= \frac{L}{2\cV_{OM}}\Delta y(\mathsf{t}) \; ,
\eeqa
between adjacent M5-branes grows, whereas the Calabi-Yau volume
and the orbifold size stay approximately constant (see
fig.\ref{Fig3}). Once, however, the M5-brane distances have grown
to a size comparable to the orbifold size $L$ itself, the $W_{99}$
open membrane instanton contribution will become of the same size
as $W_{55}$. This additional exponential contribution to the
potential will then cause the orbifold size to grow as well. This
growth will however soon end. The reason is that when the orbifold
size grows the hidden boundary Calabi-Yau volume shrinks,
rendering the hidden gauge theory strongly coupled and setting off
gaugino condensation. Once gaugino condensation is present it will
counterbalance the expansion caused by the open membrane
instantons and stabilize the orbifold modulus $T$ as worked out in
detail in \cite{BCK}.

Let us finally comment on $W_{59}, W_{95}$. Their biggest
contribution will come from the two outermost M5-branes closest to
the hidden resp.~the visible boundary. Let us estimate when their
contribution will equal $W_{55}$. Due to the symmetry of the
problem it is enough to focus on the visible boundary side. With
$|W_{95}|\simeq |h| e^{-y_i/2}$ and $|W_{55}| = |h| (N-1)
e^{-\Delta y/2}$ $\simeq |h| N e^{-\Delta y/2}$, setting both
contributions equal, gives
\beqa
y_i = \Delta y - 2\ln N \;.
\eeqa
This is equivalent to $x_i^{11}/L = \Delta x/L  - (\ln N) /
\cV_{OM}>0$, where the last inequality guarantees a positive
$x_i^{11}/L > 0$. With $N\gtrsim 36$ (coming from the spectral
index constraint) and a value $\cV_{OM}=7$ as above, this says
that only if $\Delta x/L$ has grown to surpass
\beqa
\Delta x/L > 0.5 \; ,
\eeqa
will the $W_{59}$ (and by symmetry also the $W_{95}$) become of
similar size to $W_{55}$ and need to be considered. This is,
however, close to the end of inflation and will therefore be
neglected during the inflationary phase itself.

As an indicator for when inflation comes to an end, we can
therefore use the distance between adjacent M5-branes. While at
the start of inflation at time $\mathsf{t_i}$, we have
\beqa
\frac{\Delta x(\mathsf{t_i})}{L} \ll 1
\qquad \Leftrightarrow \qquad
\Delta y(\mathsf{t_i}) \ll t \; ,
\label{Crit1}
\eeqa
we find that inflation stops at a time $\mathsf{t_f}$, when (see
fig.\ref{Fig3})
\beqa
\frac{\Delta x(\mathsf{t_f})}{L} \gtrsim \frac{1}{2}
\qquad \Leftrightarrow \qquad
\Delta y(\mathsf{t_f}) \gtrsim \frac{t}{2} \; .
\label{Crit2}
\eeqa
The reheating will happen when the M5-branes coalesce with the
visible boundary through small instanton transitions while an
additional contribution to the reheating will come from the
stabilization of $T$ towards the end of inflation. Small instanton
phase transitions were initially discovered in \cite{SITW} and
studied in connection to heterotic M-theory in \cite{SITO} (see
also \cite{GLP}). It was found that when an M5-brane disappears
into the boundary and generates a singular torsion free sheaf,
this sheaf which is referred to as a small instanton, can be
smoothed out to a smooth holomorphic vector bundle by moving in
its moduli space. This process changes the boundary's instanton
vacuum. This change of topological data will generically alter the
boundary's unbroken gauge group. For very specific initial
topological data also chirality changes can be induced, changing
the number of quark and lepton families on the visible boundary.
In contrast to the SO(32) small instanton, which can be described
in terms of some massless fields that appear at the singularity,
it is believed that a non-trivial six-dimensional conformal field
theory governs the $E_8\times E_8$ small instanton
singularity\footnote{We thank
E.~Witten for helpful comments.} \cite{SW}. It would be
interesting to explore whether cosmic strings as gauge theory
solitons or the M-theory equivalent of the recently found cosmic
superstrings of \cite{CMP} could arise in this phase transition.

But let us now, after having given a criterium for the end of
inflation, determine the number of e-foldings generated during
inflation. This number is given by
\beqa
{\mathsf N}_e \equiv \ln
\bigg(\frac{a(\mathsf{t_f})}{a(\mathsf{t_i})}\bigg)
= p \ln\Big(\frac{\mathsf{t_f}}{\mathsf{t_i}}\Big) \; ,
\eeqa
which is usually assumed to lie between 50 and 60. Since the
criterium (\ref{Crit2}) for the end of inflation is expressed in
terms of the difference $\Delta x(\mathsf{t})$, we have to express
cosmic time $\mathsf{t}$ in terms of $\Delta x(\mathsf{t})$. This
is easily done using the explicit solution for
$\varphi(\mathsf{t})$ given in (\ref{PL3}). The result is
\beqa
\frac{\mathsf{t}}{M_{Pl}} \sqrt{\frac{U_0}{p(3p-1)}}
= e^{\frac{t\Delta x(\mathsf{t})}{2L}} \; .
\eeqa
The number of e-foldings can thus be expressed in terms of the
geometrical M5-brane position difference as follows
\beqa
{\mathsf N}_e = \frac{tp}{2}\Big(\frac{\Delta
x(\mathsf{t_f})}{L}-\frac{\Delta x(\mathsf{t_i})}{L}\Big)
\simeq \frac{tp}{2} \frac{\Delta
x(\mathsf{t_f})}{L} \gtrsim \frac{tp}{4} \; .
\eeqa
The second approximation uses the fact that, at the beginning of
inflation, $\Delta x$ was much smaller than at the end
(\ref{Crit1}), while the last approximation uses (\ref{Crit2}).

To determine an actual value for ${\mathsf N}_e$, let us adopt the
same values for $\cV,\cV_{OM}$ as before. Then by using
(\ref{PLIp}) to express $p$ in terms of $N$, we arrive at
\beqa
{\mathsf N}_e \simeq \bigg(\frac{N}{12.7}\bigg)^3 \; .
\eeqa
With $N\simeq 89$, as required by the spectral index constraint,
we obtain ${\mathsf N}_e \simeq 345$. To comply with observation
one needs at least 50-60 e-foldings. There is no upper bound on
this number as everything what happens before the last 50-60
e-foldings will not be observable\footnote{We thank R.~Kallosh and
A.~Linde for helpful comments.}. We can therefore conclude that
both the spectral index and the number of e-foldings can be
obtained in a realistic regime from our proposed mechanism for
M-theory inflation.

\bigskip
\noindent {\large \bf Acknowledgements}\\[2ex]
It is a pleasure to thank V.~Balasubramanian, R.~Brandenberger,
G.~Curio, S.~Kachru, R.~Kallosh, A.~Linde, R.~Myers, B.~Ovrut,
J.~Polchinski and E.~Witten for useful discussions. The work of
K.B.~was supported in part by NSF grant PHY-0244722 and an Alfred
P.~Sloan fellowship. The work of M.B.~was supported by NSF PHY-0354401 and
an Alfred P.~Sloan fellowship. The work of A.K.~has been supported
by NSF grant PHY-0354401.

\begin{appendix}

\section{Conversion Formula}

Let us here derive the conversion from the set of M5-brane
position fields $\phi_i$ to the set of center-of-mass and
difference fields $\phi_{cm}, \Delta\phi$. The center-of-mass
field is defined as
\beqa
\phi_{cm} = \frac{1}{N}(\phi_1+\hdots+\phi_N) \; .
\eeqa
Since the differences between all neighboring M5-branes are the
same, i.e.~$\phi_{i+1,i}=\Delta\phi$, we have for the individual
M5-brane position fields the obvious expression
\beqa
\phi_i = (i-1)\Delta\phi+\phi_1 \; .
\label{a2}
\eeqa
It remains to express $\phi_1$ as a function of $\phi_{cm}$ and
$\Delta\phi$. This can be easily achieved by using
\beqa
N\phi_{cm} = \sum_{i=1}^N \phi_i
= \Delta\phi\sum_{i=0}^{N-1} i + N\phi_1
= \frac{N(N-1)}{2}\Delta\phi + N\phi_1 \; , \notag
\eeqa
from which we obtain the desired result
\beqa
\phi_1 = \phi_{cm} - \frac{N-1}{2}\Delta\phi \; .
\eeqa
Substituting this into (\ref{a2}), gives us finally the conversion
from the position fields $\phi_i$ to the center-of-mass and
difference fields $\phi_{cm},\Delta\phi$
\beqa
\phi_i = \phi_{cm}+\Big(i-\frac{N+1}{2}\Big)\Delta\phi \; .
\eeqa

\end{appendix}

 \newcommand{\zpc}[3]{{\em Z. Phys.} {\bf C\,#1} (#2) #3}
 \newcommand{\npb}[3]{{\em Nucl. Phys.} {\bf B\,#1} (#2) #3}
 \newcommand{\plb}[3]{{\em Phys. Lett.} {\bf B\,#1} (#2) #3}
 \newcommand{\prd}[3]{{\em Phys. Rev.} {\bf D\,#1} (#2) #3}
 \newcommand{\prb}[3]{{\em Phys. Rev.} {\bf B\,#1} (#2) #3}
 \newcommand{\pr}[3]{{\em Phys. Rev.} {\bf #1} (#2) #3}
 \newcommand{\prl}[3]{{\em Phys. Rev. Lett.} {\bf #1} (#2) #3}
 \newcommand{\jhep}[3]{{\em JHEP} {\bf #1} (#2) #3}
 \newcommand{\jcap}[3]{{\em JCAP} {\bf #1} (#2) #3}
 \newcommand{\cqg}[3]{{\em Class. Quant. Grav.} {\bf #1} (#2) #3}
 \newcommand{\prep}[3]{{\em Phys. Rep.} {\bf #1} (#2) #3}
 \newcommand{\fp}[3]{{\em Fortschr. Phys.} {\bf #1} (#2) #3}
 \newcommand{\nc}[3]{{\em Nuovo Cimento} {\bf #1} (#2) #3}
 \newcommand{\nca}[3]{{\em Nuovo Cimento} {\bf A\,#1} (#2) #3}
 \newcommand{\lnc}[3]{{\em Lett. Nuovo Cimento} {\bf #1} (#2) #3}
 \newcommand{\pra}[3]{{\em Pramana} {\bf #1} (#2) #3}
 \newcommand{\ijmpa}[3]{{\em Int. J. Mod. Phys.} {\bf A\,#1} (#2) #3}
 \newcommand{\rmp}[3]{{\em Rev. Mod. Phys.} {\bf #1} (#2) #3}
 \newcommand{\ptp}[3]{{\em Prog. Theor. Phys.} {\bf #1} (#2) #3}
 \newcommand{\sjnp}[3]{{\em Sov. J. Nucl. Phys.} {\bf #1} (#2) #3}
 \newcommand{\sjpn}[3]{{\em Sov. J. Particles \& Nuclei} {\bf #1} (#2) #3}
 \newcommand{\splir}[3]{{\em Sov. Phys. Leb. Inst. Rep.} {\bf #1} (#2) #3}
 \newcommand{\tmf}[3]{{\em Teor. Mat. Fiz.} {\bf #1} (#2) #3}
 \newcommand{\jcp}[3]{{\em J. Comp. Phys.} {\bf #1} (#2) #3}
 \newcommand{\cpc}[3]{{\em Comp. Phys. Commun.} {\bf #1} (#2) #3}
 \newcommand{\mpla}[3]{{\em Mod. Phys. Lett.} {\bf A\,#1} (#2) #3}
 \newcommand{\cmp}[3]{{\em Comm. Math. Phys.} {\bf #1} (#2) #3}
 \newcommand{\jmp}[3]{{\em J. Math. Phys.} {\bf #1} (#2) #3}
 \newcommand{\pa}[3]{{\em Physica} {\bf A\,#1} (#2) #3}
 \newcommand{\nim}[3]{{\em Nucl. Instr. Meth.} {\bf #1} (#2) #3}
 \newcommand{\el}[3]{{\em Europhysics Letters} {\bf #1} (#2) #3}
 \newcommand{\aop}[3]{{\em Ann. of Phys.} {\bf #1} (#2) #3}
 \newcommand{\jetp}[3]{{\em JETP} {\bf #1} (#2) #3}
 \newcommand{\jetpl}[3]{{\em JETP Lett.} {\bf #1} (#2) #3}
 \newcommand{\acpp}[3]{{\em Acta Physica Polonica} {\bf #1} (#2) #3}
 \newcommand{\sci}[3]{{\em Science} {\bf #1} (#2) #3}

\bibliographystyle{plain}

\end{document}